\documentclass{ws-procs9x6}
\usepackage{mcite}
\usepackage{url}
\usepackage{epsfig}

\def\msbar{\ensuremath{{\rm{\overline{MS}}}}}

\newcommand{\gev    }{\ensuremath{\mathrm{GeV}}}
\newcommand{\gevsq  }{\ensuremath{\mathrm{GeV^2}}}
\newcommand{\Ord}{\ensuremath{{O}}}

\begin{document}

\title{Heavy-flavor photo- and electroproduction at NLO
}

\author{I.~SCHIENBEIN 
}

\address{Southern Methodist University \\
Department of Physics \\ 
104 Fondren Science Building \\
3215 Daniel Avenue \\ 
Dallas, Texas 75275-0175, USA\\
E-mail: schien@physics.smu.edu}


\maketitle
                      
\abstracts{
We review one-particle inclusive production of 
heavy-flavored hadrons in a framework
which resums the large collinear logarithms through the evolution 
of the FFs and PDFs and, at the same time,
retains the full dependence on the heavy-quark mass without 
additional theoretical assumptions.
The main focus is on 
the production of $D^\star$ mesons in deep inelastic 
electron--proton scattering at HERA.
We show results, neglecting for the time being the
heavy-quark mass terms,
for deep inelastic $D^\star$ meson production at 
finite transverse momenta.
Work to implement this process in the above mentioned
massive QCD framework is in progress.
}

\section{Introduction}
One-particle inclusive production processes provide 
extensive tests of perturbative quantum chromodynamics (pQCD).
In contrast to fully inclusive processes, it is possible to study
distributions in the momentum of the final state particle and 
to apply kinematical cuts which come close to the experimental 
situation.
On the other hand, contrary to even more exclusive cases,
QCD factorization theorems \cite{Collins:1989gx,Collins:1987pm}
still hold stating that this
class of observables can be computed as convolutions of
{\em universal} parton distribution functions (PDFs) and
fragmentation functions (FFs) with perturbatively 
calculable hard scattering cross sections.
As is well-known, it is due to the factorization property that
the parton model of QCD has predictive power.
Hence, tests of the universality of the PDFs and FFs are of
crucial importance for validating this QCD framework.
At the same time, lowest order expressions for the hard scattering
cross sections are often not sufficient for meaningful tests 
and the use of higher order computations is mandatory.

The perturbative analysis is becoming more involved and interesting
if the observed final state hadron contains a heavy (charm or bottom)
quark. In this case, the heavy-quark mass $m_h$ enters as an 
additional scale.
Clearly, the conventional massless formalism, also known
as zero-mass variable-flavor-number scheme (ZM-VFNS), can also be
applied to this case, provided the hard scale $Q$ of the process
is much bigger than the heavy-quark mass 
such that terms
$m_h/Q$ are negligible.
However, at present collider energies, most of the experimental data
lie in the kinematic region $Q \gtrsim m_h$ and it is necessary
to take the power-like mass terms into account in a consistent
framework.

The subject of this article is the 
theoretical
description of 
one-particle inclusive production of heavy-flavored 
hadrons $X_h=D,B,\Lambda_c,\ldots$
in a massive variable-flavor-number scheme (GM-VFNS).
In such a scheme the large collinear logarithms of the 
heavy-quark mass $\ln \mu/m_h$ 
are subtracted from the hard scattering cross sections and
resummed through the evolution of the fragmentation functions (FFs)
and parton distribution functions (PDFs).
At the same time, finite non-logarithmic mass terms $m_h/Q$
are kept in the hard part and 
fully taken into account.

In order to test the pQCD formalism, in particular the
universality of the FFs, it is important to provide a 
description of all relevant processes in a coherent framework.
Therefore,
it is important to work out the GM-VFNS at
next-to-leading order (NLO) of QCD
for all the relevant processes.
Previously, the GM-VFNS has been applied to the following
processes:
$\gamma + \gamma \to D^{\star +} + X$ (direct part) \cite{Kramer:2001gd},
$\gamma + \gamma \to D^{\star +} + X$ (single resolved part) 
\cite{Kramer:2003cw},
$\gamma + p \to D^{\star +} + X$ (direct part) 
\cite{Kramer:2003jw},
$p + \bar{p} \to (D^0,D^{\star +},D^+,D_s^+) + X$
\cite{Kniehl:2004fy,Kniehl:2005mk,Kniehl:2005ej},
where the latter results for hadron--hadron collisions 
also constitute the resolved contribution
to the photoproduction process $\gamma + p \to X_h + X$.

In this contribution, 
we will review the production of heavy-flavored hadrons $X_h$
in hadron--hadron, photon--proton and deep inelastic 
electron--proton collisions 
where the main focus will be on the electroproduction case.
We will show results, neglecting for the time being the $m_h/Q$ mass 
terms, for deep inelastic $D^\star$ meson production at 
finite transverse momenta.

\section{Theoretical Framework}
\subsection{GM-VFNS}
The differential cross sections for inclusive heavy-flavored hadron
production can be computed in the GM-VFNS according to the
familiar factorization formulae, 
however, with heavy-quark mass terms included in the hard scattering
cross sections \cite{Collins:1998rz}.
Generically, the physical cross sections are expressed as convolutions
of PDFs for the incoming hadron(s), hard scattering cross sections, and
FFs for the fragmentation of the outgoing parton into the observed
hadron. All possible partonic subprocesses are taken into account.
The massive hard scattering cross sections are constructed in a way
that in the limit $m_h \to 0$ the conventional ZM-VFNS is recovered.
A more detailed discussion of the GM-VFNS and the construction of
the massive hard scattering cross sections can be found
in Refs.\ \refcite{Kniehl:2004fy,Kniehl:2005mk}
and the conference proceedings 
\cite{Schienbein:2003et,*Schienbein:2004ah,*Kniehl:2005st,Baines:2006uw}.
\subsection{Fragmentation Functions}
A crucial ingredient entering these calculation are the 
non-perturbative FFs for the transition
of the final state parton into the observed hadron $X_h$.
For charm-flavored mesons, $X_c$, three sets of FFs have been
employed in our analyses:
(i) BKK98-D \cite{Binnewies:1998xq}:
For $X_c=D^{*+}$, such FFs were extracted at LO and NLO in the
$\overline{\rm MS}$ factorization scheme with $n_f=5$ massless quark flavors
several years ago \cite{Binnewies:1998xq} from the 
scaled-energy ($x$) distribution
$d\sigma/dx$ of the cross section of $e^+e^-\to D^{*+}+X$ measured by the
ALEPH \cite{Barate:1999bg} and OPAL \cite{Ackerstaff:1997ki} 
Collaborations at CERN LEP1.
(ii) KK05-D \cite{Kniehl:2005de}:
Recently, the BKK98-D analysis was extended \cite{Kniehl:2005de}
to include $X_c=D^0,D^+,D_s^+,\Lambda_c^+$ by exploiting
appropriate OPAL data \cite{Alexander:1996wy}.
(iii) KK05-D2:
In Refs.~\cite{Binnewies:1998xq,Kniehl:2005de}, the starting scales 
$\mu_0$ for the DGLAP evolution of the $a\to X_c$ FFs in the 
factorization scale $\mu_F^\prime$ have been
taken to be $\mu_0=2 m_c$ for
$a=g,u,\overline{u},d,\overline{d},s,\overline{s},c,\overline{c}$ and
$\mu_0=2 m_b$, with $m_b=5$~GeV, for $a=b,\overline{b}$.
The FFs for $a=g,u,\overline{u},d,\overline{d},s,\overline{s}$ were assumed to
be zero at $\mu_F^\prime=\mu_0$ and were generated through the DGLAP evolution
to larger values of $\mu_F^\prime$. 
For consistency with the $\msbar$ prescription for PDFs, we repeated
the fits of the $X_c$ FFs for the choice $\mu_0 = m_c,m_b$.
This changes the $c$-quark FFs only marginally, but has an 
appreciable effect on the gluon FF, which is
important at Tevatron energies, as was
found for $D^{*+}$ production in Ref.~\cite{Kniehl:2004fy}.
These new FFs will be presented elsewhere.

\subsection{Input Parameters}
\label{sec:parameters}
If not stated otherwise, the following parameters 
have been chosen for the numerical results presented
below.
For the proton PDFs we have employed the CTEQ6M/CTEQ6.1M PDFs 
from the CTEQ Collaboration
\cite{Pumplin:2002vw,*Stump:2003yu}
and the KK05-D2 FFs.
We have set $m_c = 1.5\ \gev$, $m_b = 5\ \gev$ and have used 
the two-loop formula for $\alpha_s^{(n_f)}(\mu_R)$ in the
$\msbar\ $ scheme with $\alpha_s^{(5)}(m_Z) = 0.118$.
Our theoretical predictions depend on three scales,
the renormalization scale $\mu_R$, and the initial- and final-state
factorization scales $\mu_F$ and $\mu_F^\prime$, respectively.
Our default choice for hadro- and photoproduction has been
$\mu_R = \mu_F = \mu_F^\prime = m_T$, where $m_T = \sqrt{p_T^2 + m_h^2}$
is the transverse mass.
The scale choice in the electroproduction case will be specified below.

\section{Hadroproduction}
Recently, the CDF collaboration has published first cross section data
for inclusive  production of $D^0$, $D^+$, $D^{*+}$, and $D_s^+$ mesons
in $p\bar{p}$ collisions \cite{Acosta:2003ax} obtained in Run II at
the Tevatron at center-of-mass energies of $\sqrt{S} = 1.96$ TeV.
The data come as distributions $d\sigma/dp_T$ with $y$ integrated over 
the range $|y|\le1$ and the particle and antiparticle contributions are
averaged.

\begin{figure*}[t]
\begin{center}
\begin{tabular}{ll}
{\parbox{6.0cm}{
\epsfig{file=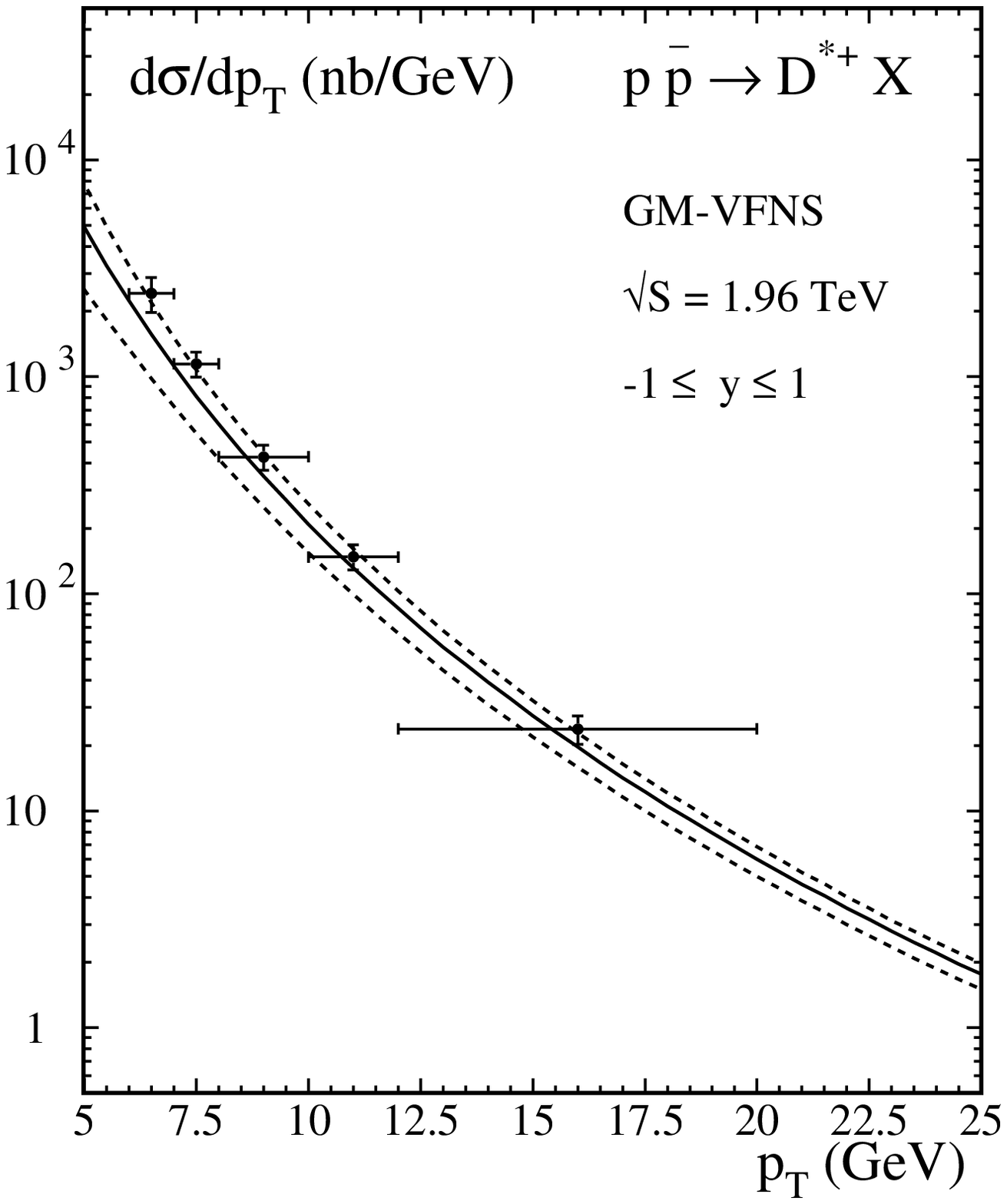,width=5.5cm}
}} &
{\parbox{6.0cm}{
\hspace*{-1.5cm}
\epsfig{file=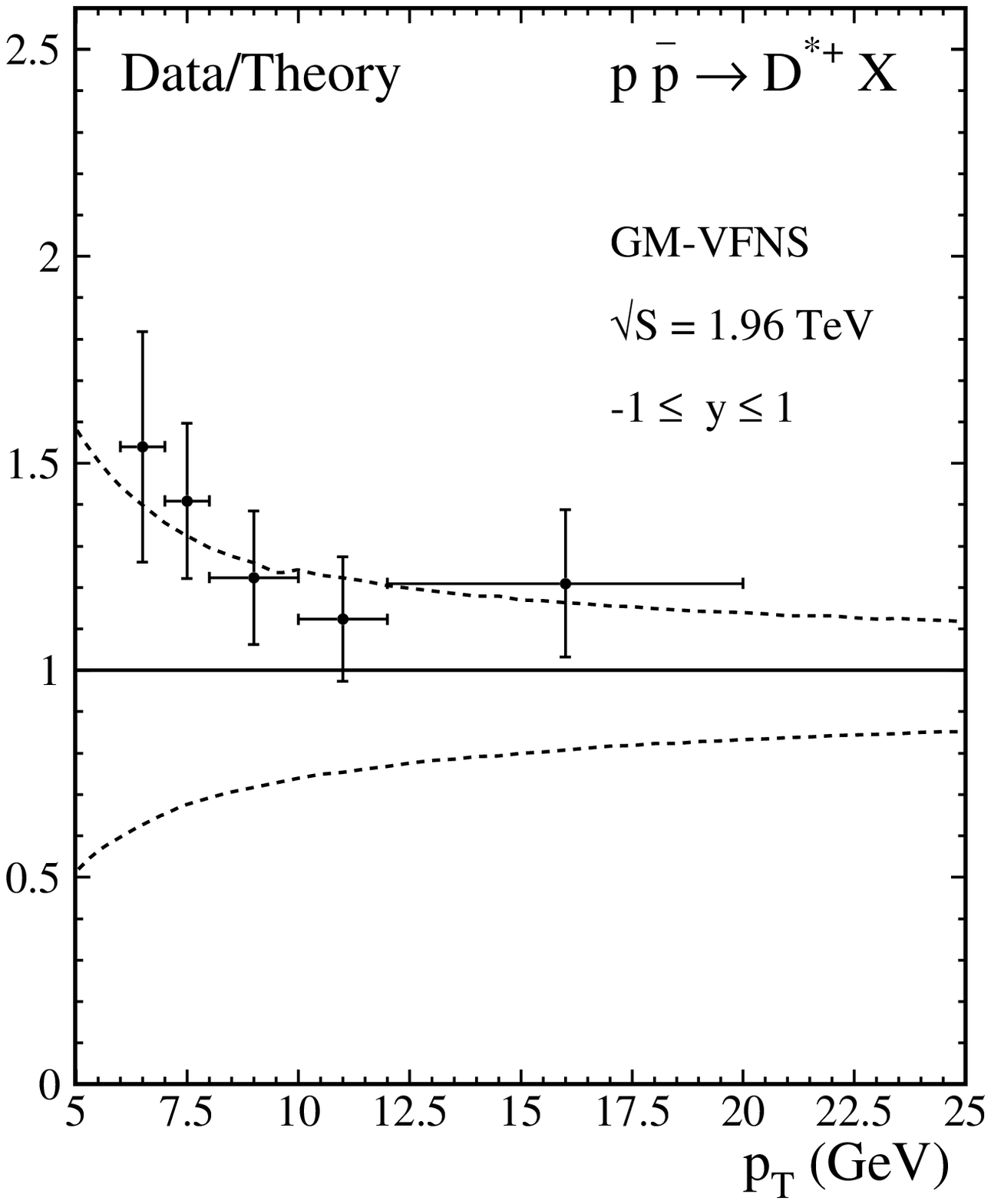,width=5.5cm}
}}
\end{tabular}
\end{center}
\caption{
Comparison of the CDF data \protect\cite{Acosta:2003ax} 
with our NLO predictions for
$D^{*+}$.
The solid line represents our default prediction
obtained with $\mu_R = \mu_F = \mu_F^\prime = m_T$,
while the dashed lines
indicate the scale uncertainty 
estimated by varying $\mu_R$, $\mu_F$, and $\mu_F^\prime$ 
independently within a factor of 2 up and down relative to 
the central values.
The right figure shows
the data-over-theory representation 
with respect to our default prediction.
}
\label{fig:fig1}
\end{figure*}

Our theoretical predictions in the GM-VFNS 
are compared with the CDF data for $D^\star$ mesons on an absolute
scale in Fig.~\ref{fig:fig1} (left) and in the data-over-theory representation
with respect to our default results in Fig.~\ref{fig:fig1} (right).
We find good agreement in the sense that the theoretical and
experimental errors overlap where
the experimental results are gathered
on the upper side of the theoretical error band, corresponding to a small
value of $\mu_R$ and large values of $\mu_F$ and $\mu_F^\prime$, the $\mu_R$
dependence being dominant in the upper $p_T$ range.
As is evident from Fig.~\ref{fig:fig1} (right), the central data
points tend to overshoot the central QCD prediction by a factor of about 1.5
at the lower end of the considered $p_T$ range, where the errors are largest,
however.
This factor is rapidly approaching unity as the value of $p_T$ is increased.
The tendency of measurements of inclusive hadroproduction in Tevatron run~II
to prefer smaller renormalization scales is familiar from single jets, which
actually favor $\mu_R=p_T/2$~\cite{field05}.

For more details and a comparison 
with the data for the $D^0$, $D^+$, and $D_s^+$ mesons
we refer to Ref.\ \refcite{Kniehl:2005ej}.
A comparison of NLO predictions 
for $p + \bar{p} \to B^+ +X$ in the GM-VFNS 
with recent CDF data \protect\cite{Acosta:2004yw} is in 
preparation \cite{winp}.

\section{Photoproduction}
Inclusive photoproduction of $D^\star$ mesons, $\gamma + p \to D^\star + X$, 
has been studied in Ref.\ \refcite{Kramer:2003jw} where the direct part 
has been computed in the GM-VFNS whereas the resolved part has been 
included in the ZM-VFNS.
In this analysis the BKK98-D FFs have been utilized and, for the resolved
contribution, the GRV92 photon PDFs \cite{cit:GRV-9202}.
The other parameters have been chosen as specified 
in Sec.\ \ref{sec:parameters}.
In Fig.\ 6 of Ref.\ \refcite{Kramer:2003jw}, the central numerical
predictions for the transverse momentum ($p_T$)
distributions of the $D^\star$ meson have been compared
with preliminary ZEUS data \cite{Zeus:photo}.
There exist similar data by the H1 collaboration
\cite{H1:photo} which have not been used in this analysis.
As can be seen in this figure, the agreement of the 
$p_T$-distributions with the data is quite good down to
$p_T \simeq 2 m_c$ and the mass effects turn out to be small.
In order to extend the range of applicability of the GM-VFNS
into the region $p_T < 3\ \gev$ more work on the matching to
the 3-fixed flavor theory would be needed.
The Figs.\ 7 -- 9 of  Ref.\ \refcite{Kramer:2003jw},
showing results for the rapidity ($y$), invariant mass ($W$) and
inelasticity ($z(D^\star)$) distributions, have to be taken with a
grain of salt since they receive large contributions from the 
transverse momentum region $1.9 < p_T < 3\ \gev$ which is outside 
the range of validity of the present theory.

With the work in Ref.\ \refcite{Kniehl:2004fy},
it is now possible to include also the resolved part in the GM-VFNS.
It will be interesting to compare
the complete GM-VFNS framework at NLO of QCD, combined with 
updated FFs, in more detail with ZEUS and H1 photoproduction data 
once they are finalized.

\section{Electroproduction}
Recently, the single inclusive electroproduction of light hadrons
at finite transverse momenta has attracted quite a lot of interest,
where the outgoing hadron is required to carry a non-vanishing
transverse momentum ($p_T^*$) in the center-of-mass system (CMS)
of the virtual photon and the incoming proton.
The following partonic subprocesses contribute at leading order (LO)
and are of the order $\Ord(\alpha_s)$:
$\gamma^\star + q \to q + g$ and $\gamma^\star + g \to q + \bar{q}$.
Very recently, the NLO ($\Ord(\alpha_s^2)$) corrections to this 
process have been 
accomplished by three
independent groups \cite{Aurenche:2003by,Daleo:2004pn,Kniehl:2004hf}.
Suffice to say here that the NLO corrections increase the LO results
in certain kinematical regions by large factors
and are essential for bringing theory in agreement with the 
experimental results.
For more details see Ref.\ \refcite{Bernd05}. 

\begin{figure*}[t]
\begin{center}
\epsfig{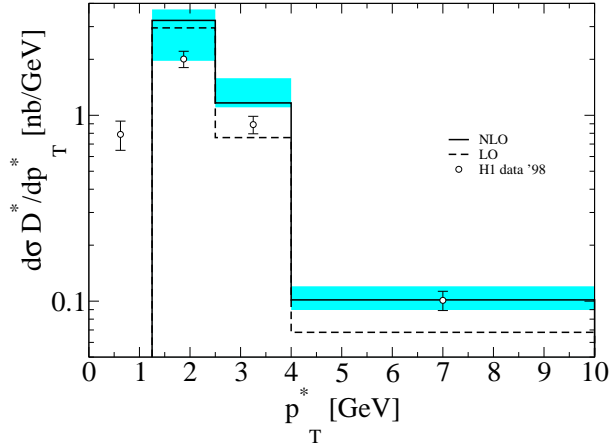}
\end{center}
\caption{
NLO QCD predictions in the ZM-VFNS for electroproduction
of $D^\star$ mesons in dependence of the transverse momentum of
the $D^\star$ meson in the $\gamma^\star p$ CMS.
H1 data \protect\cite{Adloff:1998vb} are shown for comparison.
See the text for further details.}
\label{fig:fig2}
\end{figure*}

\begin{figure*}[t]
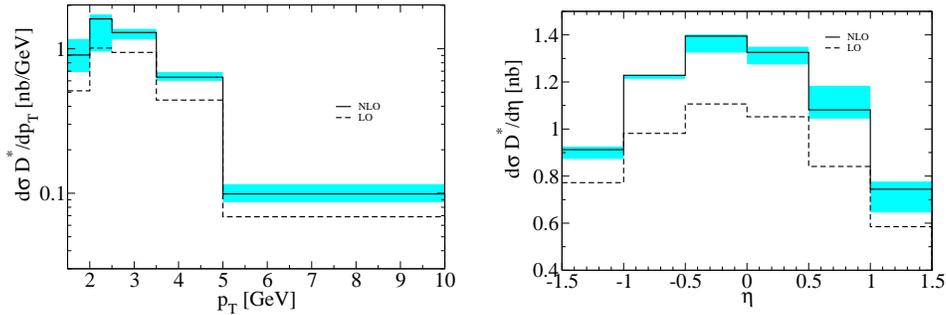

\begin{center}
\begin{tabular}{ll}
{\parbox{5.0cm}{
\hspace*{-1.2cm}
\epsfig{file=schien_fig3a.eps,width=6.0cm}
}} &
{\parbox{5.0cm}{
\epsfig{file=schien_fig3b.eps,width=6.0cm}
}} 
\end{tabular}
\end{center}
\caption{
NLO QCD predictions in the ZM-VFNS for electroproduction
of $D^\star$ mesons in dependence of the transverse momentum (left)
and the pseudo-rapidity (right) of the $D^\star$ meson in the 
laboratory frame.
For further details on the kinematical cuts and the uncertainty band
see the text.
}
\label{fig:fig3}
\end{figure*}

\begin{figure*}[t]
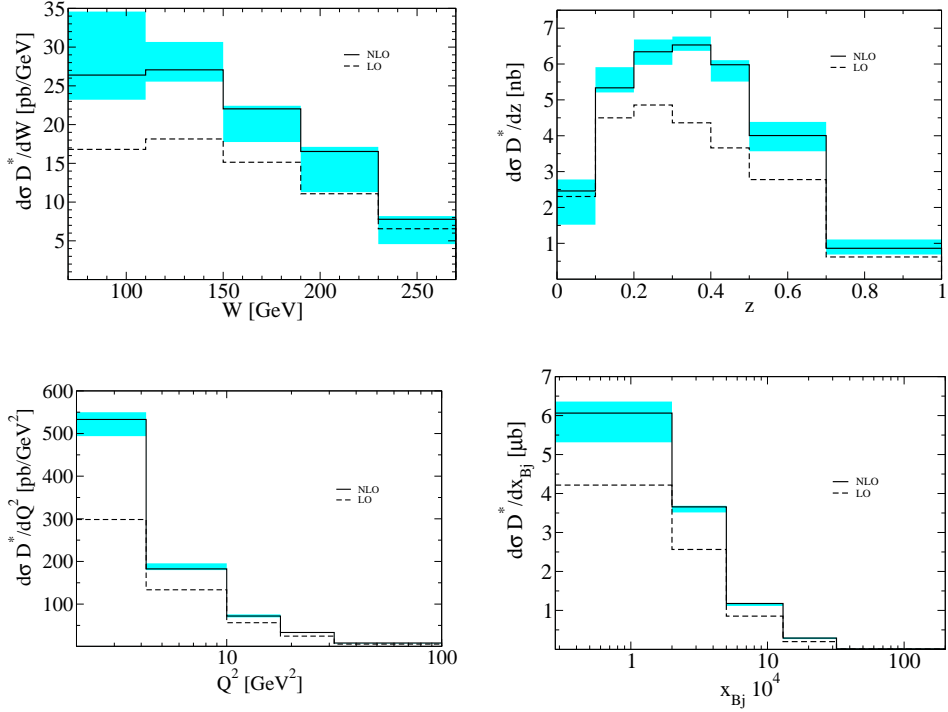

\begin{center}
\begin{tabular}{ll}
{\parbox{5.0cm}{
\hspace*{-1.2cm}
\epsfig{file=schien_fig4a.eps,width=6.0cm}
}} &
{\parbox{5.0cm}{
\epsfig{file=schien_fig4b.eps,width=6.0cm}
}} 
\vspace*{0.6cm}
\\
{\parbox{5.0cm}{
\hspace*{-1.2cm}
\epsfig{file=schien_fig4c.eps,width=6.0cm}
}} &
{\parbox{5.0cm}{
\epsfig{file=schien_fig4d.eps,width=6.0cm}
}} 
\end{tabular}
\end{center}
\caption{
As in Fig.\ \protect\ref{fig:fig3} for the $\gamma^\star p$ 
invariant mass $W$, the inelasticity $z$, the momentum transfer $Q^2$ 
and Bjorken-$x$.
}
\label{fig:fig4}
\end{figure*}

Endowed with appropriate fragmentation functions, 
the computation in Ref.\ \refcite{Kniehl:2004hf} has been 
employed to obtain predictions in the ZM-VFNS for the production 
of heavy-flavored hadrons in electron--proton collisions at 
HERA, $e + p \to e + X_h + X$.
The electron and proton energies have been set to $E_e = 27.5\ \gev$ and
$E_p = 820\ \gev$ (Fig.\ \ref{fig:fig2}) resp.\ $920\ \gev$ 
(Figs.\ \ref{fig:fig3}, \ref{fig:fig4}) in the laboratory frame.
Furthermore, the renormalization and factorization scales have been
chosen \cite{Kniehl:2004hf} as 
$\mu_R^2 = \mu_F^2 = {\mu_F^\prime}^2 = \xi \frac{Q^2 + (p_T^*)^2}{2}$.
The dimensionless parameter $\xi$ has been varied between $1/2$ and
$2$ about the default value $1$ in order to estimate the scale
uncertainties of the theoretical predictions.
The other input parameters have been chosen as described in 
Sec.\ \ref{sec:parameters}.
The following cuts have been imposed on the numerical results
in Figs.\ \ref{fig:fig2}--\ref{fig:fig4}:
$2 < Q^2 < 100\ \gevsq$, $0.05 < y < 0.7$,
$1.5 \le p_{T,{\rm Lab}}(D^\star) \le 15\ \gev$,
and
$|\eta_{\rm Lab}(D^\star)| < 1.5$
where the momentum transfer $Q^2$, the inelasticity $y$, and the 
pseudorapidity $\eta$ are defined as usual.
Furthermore, in Figs.\ \ref{fig:fig3}--\ref{fig:fig4} we have
asked for the additional constraint $p_T^*(D^\star) > 2\ \gev$
where $p_T^*(D^\star)$ is the transverse momentum of the 
$D^\star$ meson in the $\gamma^* p$-CMS.
This cut is essential to avoid the collinear singularities 
as $p_T^* \to 0$. The results have been calculated with
$n_f = 5$ flavors and include the contribution where a bottom
quark fragments into the $D^\star$ meson via $b \to B \to D^\star$.
No distinction is made between $D^{\star +}$ and $D^{\star -}$.
Figure \ref{fig:fig2} shows the $p_T^*$ spectrum in comparison with 
H1 data \cite{Adloff:1998vb}, collected in the years 1994 to 1996
with 27.5 $\gev$ positrons colliding with 820 $\gev$ protons
at CMS energies of $\sqrt{S} = 300\ \gev$.
The data points are reasonably well described 
if one keeps in mind that the theory is expected to work for
larger $p_T^* \gtrsim 2\ \gev$.
Furthermore, in the ZM-VFNS the heavy-quark mass terms are missing
which are potentially important in the region of small $p_T^*$.
These terms will be included as soon as ongoing work
to implement this process in the
GM-VFNS has been completed.

Finally, 
results for other distributions ($p_{T,{\rm Lab}}$, $\eta_{\rm Lab}$,
$W$, $z$, $Q^2$, and $x_{\rm Bj}$) are presented in
Figs.\ \ref{fig:fig3}--\ref{fig:fig4}.

\section{Summary}
We have discussed one-particle inclusive production of heavy-flavored
hadrons in hadron--hadron, photon--proton, and electron--proton collisions
in a massive variable-flavor-number scheme (GM-VFNS).
The importance of a unified treatment of all these processes, based on
QCD factorization theorems, has been emphasized, in order to provide
meaningful tests of the universality of the FFs and hence of QCD.
At the same time, it is necessary to incorporate heavy-quark mass effects
in the formalism since many of the present experimental data points lie in
a kinematical region where the hard scale of the process is not much
larger than the heavy-quark mass.
This is achieved in the GM-VFNS, which includes heavy-quark mass effects
in a rigorous way and still relies on QCD factorization.
We have discussed numerical results for the three reactions.
In general, the description of the transverse momentum spectra is
quite good down to transverse momenta $p_T \simeq 2 m_h$.
Extending the range of applicability of our scheme to smaller
$p_T$ would require more work on the matching to the corresponding
theories in the fixed flavor number schemes.
Our ZM-VFNS results for the electroproduction of $D^\star$ mesons 
indicate that future experimental
results by the H1 collaboration can be nicely described.
All this leads us to the expectation that a good overall 
description of the data can be reached 
in the future

\section*{Acknowledgments}
The author would like to thank the organizers 
of the Ringberg workshop 
on {\it New Trends in HERA Physics 2005}
for the kind invitation, 
B.\ A.\ Kniehl, G.\ Kramer and H.\ Spiesberger 
for their collaboration, and
M.\ Maniatis for providing the figures
\ref{fig:fig2} -- \ref{fig:fig4} and for useful discussions.

\end{document}